\documentclass[pdflatex,sn-nature]{sn-jnl}

\usepackage{graphicx}%
\usepackage{multirow}%
\usepackage{amsmath,amssymb,amsfonts}%
\usepackage{amsthm}%
\usepackage{mathrsfs}%
\usepackage[title]{appendix}%
\usepackage{xcolor}%
\usepackage{textcomp}%
\usepackage{manyfoot}%
\usepackage{booktabs}%
\usepackage{algorithm}%
\usepackage{algorithmicx}%
\usepackage{algpseudocode}%
\usepackage{listings}%
\usepackage{aas_macros}%
\usepackage{subfig}

\begin{document}

\title[Tearing Mode Quenched by Transverse Fields]{Quenching of Tearing Mode Instability by  Transverse Magnetic Fields in Reconnection Current Sheets}


\author*[1]{\fnm{Grzegorz} \sur{Kowal}}\email{grzegorz.kowal@usp.br}
\equalcont{These authors contributed equally to this work.}

\author[1]{\fnm{Diego A.} \sur{Falceta-Gonçalves}}\email{dfalceta@usp.br}
\equalcont{These authors contributed equally to this work.}

\affil*[1]{\orgdiv{Escola de Artes, Ciências e Humanidades}, \orgname{University of São Paulo}, \orgaddress{\street{Av. Arlindo Béttio, 1000 - Vila Guaraciaba}, \city{São Paulo}, \postcode{03828-000}, \state{SP}, \country{Brazil}}}


\abstract{The tearing mode instability is a key process for magnetic energy conversion in magnetohydrodynamics, once anti-parallel components are allowed to reconnect, leading to the formation of magnetic islands. It has been employed to explain phenomena at different scales in nature, from galactic nuclei, to solar flares and laboratory fusion devices. In this study, we investigate the dynamics of a current sheet in the presence of a transverse magnetic field component, in the framework of viscoresistive, incompressible magnetohydrodynamics (MHD), both analytically and by means of direct numerical simulations. Firstly, we obtain analytical solution for the time-varying one-dimensional profile of an initial Harris current sheet in the presence of a transverse field. We find that the introduction of a transverse magnetic field disrupts the system's equilibrium, leading to the natural development of a neutral layer with shear flows within the current sheet, one along the antiparallel magnetic component and another along the guide field direction. Secondly, through numerical analysis, we examine the dispersion relation of the incompressible MHD equations in the context of a modified equilibrium profile due to the transverse field. Our findings indicate a rapid suppression of unstable modes of tearing instability with the width of the neutral layer, confirming the analytical predictions. These results offer new insightful understanding on the interplay between transverse magnetic fields, shear flows, and tearing mode instabilities in current sheet environments.}




\maketitle

Tearing instability is a fundamental process in magnetohydrodynamics (MHD) that plays a critical role in the dynamics of astrophysical plasmas. First analyzed in detail by Furth, Killeen, and Rosenbluth (1963) \cite{1963PhFl....6..459F}, the tearing is a type of resistive instability that disrupts current sheets in plasmas, leading to faster magnetic reconnection as compared to standard Sweet-Parker model \cite{1957JGR....62..509P, 1958IAUS....6..123S}, a process essential for converting magnetic energy into kinetic and thermal energy, as well as accelerating particles, which is significant in various astrophysical environments.

In astrophysical contexts, magnetic reconnection is crucial for understanding phenomena such as solar flares, magnetic storms, and the heating of the solar corona \cite{1993SSRv...65..253S}. These events are driven by the release of stored magnetic energy through reconnection, which is supposedly facilitated by the tearing instability. This instability enables the breaking and reconnection of magnetic field lines, leading to a rapid reconfiguration of the magnetic field topology and the associated energy release \cite{2006ApJ...645..732B}.

Recent research on the interplay between turbulence and tearing mode instability suggests a crucial role of the instability in shaping the energy dissipation and magnetic field structure in magnetohydrodynamic (MHD) turbulence \cite{2017PhRvL.118x5101L, 2017ApJ...844..125B, 2018PhRvE..98c3209W}. Current sheets formed within turbulent eddies can become unstable to tearing modes when their thickness lies below a critical scale, leading to the formation of plasmoids and altering the energy cascade. This tearing-mediated regime introduces a sub-inertial interval characterized by a distinct scaling of the energy spectrum, different from the standard Kolmogorov-like turbulence. For instance, the scale of eddies where reconnection becomes significant is larger than the resistive diffusion scale, resulting in a $k_{\perp}^{-5/2}$ spectrum in the reconnection interval. The instability also affects the alignment of magnetic field lines, reducing anisotropy at smaller scales and modifying the turbulence spectrum to $k_{\perp}^{-11/5}$. Furthermore, in the context of the turbulent dynamo, tearing instability disrupts folded magnetic structures at high magnetic Reynolds numbers, enhancing viscous-to-resistive dissipation and steepening the magnetic energy spectrum \cite{2022PhRvX..12d1027G}. These findings highlight that tearing instability not only competes with but also significantly influences the nonlinear evolution of turbulent eddies, leading to a complex and dynamic interplay that is essential for understanding energy dissipation and magnetic field topology in various astrophysical and laboratory plasma environments.

Plasmoid instability, which has become synonymous of tearing mode instability in a three-dimensional scenario, is a significant mechanism in magnetic reconnection processes. This instability manifests when thin current sheets, formed during magnetic reconnection events, become susceptible to high-wavenumber perturbations, leading to the rapid formation of fluxtubes trapped between reconnecting active regions, known as plasmoids or secondary islands. Theoretical models and numerical simulations \cite{2007PhPl...14j0703L, 2009PhPl...16k2102B, 2012PhPl...19b2101B, 2016PhPl...23j0702C}, have shown that the growth rate of plasmoid instability scales with the Lundquist number, $S=Lv_A/\eta$, following the relation $\gamma \sim S^{1/4}$. This rapid instability causes the current sheets to break into a chain of plasmoids, accelerating the reconnection process beyond the rates predicted by the classical Sweet-Parker model. In high-Lundquist-number plasmas, plasmoid instability induces a regime of fast reconnection characterized by dynamic and impulsive behaviors, leading to significantly higher reconnection rates than previously understood. The ejection of the plasmoids, however, is necessary for an efficient magnetic energy conversion - and particle acceleration -, which are crucial in explaining the rapid energetic events observed in astrophysical and laboratory plasmas.

Viscosity plays a significant role in modulating the tearing mode instability, particularly in high-Lundquist-number plasmas \cite{1990PhFlB...2.2575C, 1991PhFlB...3.1364O, 2013PhRvE..87a3102L, 2015ApJ...801..145T}. The introduction of viscosity alters the scaling of the growth rate and the behavior of the unstable modes. For instance, when viscosity is considered alongside small shear flow, the scaling of tearing growth rate changes to $S^{-2/3} (S_v/S)^{1/6}$, where $S_v$ is the ratio of the viscous time to the Alfvén time. This adjustment in scaling indicates that viscosity can strongly reduce the instability growth rate compared to the inviscid case. Additionally, at large magnetic Prandtl numbers ($ P_m = \nu/\eta \gg 1 $), the growth rate and wavenumber of the most unstable mode scale differently, leading to predictions such as $\gamma_\mathrm{max} \sim S^{1/4} P_m^{-5/8}$ and $k_\mathrm{max} L_{CS} \sim S^{3/8} P_m^{-3/16}$. This implies that higher viscosity allows for larger aspect ratios of current sheets before they become unstable. In the context of Sweet-Parker current sheets, the presence of viscosity also affects the onset and growth rate of the plasmoid instability and, in some cases, it enables the Kelvin-Helmholtz instability to grow faster than the plasmoid instability. Overall, viscosity introduces additional complexity to the tearing mode dynamics influencing the stability and structure of current sheets in both laboratory and astrophysical plasmas.

The effects of compressibility and velocity shear significantly influence the tearing mode instability in various ways. Compressibility enhances the development of tearing instabilities, particularly in regions where incompressible plasma would otherwise remain stable \cite{1993ARep...37..282V}. This destabilizing effect is triggered by the longitudinal magnetic field, playing a crucial role in dynamic solar phenomena such as flares. On the other hand, velocity shear, particularly in the presence of equilibrium shear flow and viscosity, introduces complex modifications to the tearing mode growth rate \cite{1990PhFlB...2.2575C, 1991PhFlB...3.1364O, 2018JPlPh..84a9015T, 2018ApJ...859...83S}. Shear flow alters the eigenfunction profiles, creating new peaks outside those found in flow-free scenarios, and modifies growth rate scalings to $ S^{-2/3} (S_v/S)^{1/6} $. High shear flow can induce Kelvin-Helmholtz instabilities, further complicating the dynamics. Additionally, shear flow can stabilize the current sheet by stretching and evacuating magnetic islands, limiting the duration of instability growth and leading to non-exponential behavior. The interplay between these factors—compressibility and shear flow—demands detailed numerical and analytical studies to fully understand their combined impact on magnetic reconnection processes in astrophysical and laboratory plasmas.

In a three-dimensional (3D) tearing scenario, one expects a complex interplay between multiple resonant surfaces and the influence of a guide field on the instability dynamics \cite{2012PhPl...19b2101B, 2020ApJ...902..142S}. Unlike the two-dimensional (2D) case, where instability occurs primarily at the null surface, 3D geometries with a guide field exhibit a spectrum of unstable modes across various resonant surfaces within the current sheet. These surfaces are characterized by unique angles of obliquity, with the most unstable modes being oblique rather than parallel. The presence of a guide field introduces a dispersive nature to the propagation of these modes, altering their growth rates and structures. For instance, the most unstable wavenumber is found at the intersection of constant-$\psi$ and non-constant-$\psi$ regimes (where $\psi$ is the magnetic flux function), with the growth rate scaling as $\gamma_\mathrm{max} \sim S_L^{1/4}(1-\mu^4)^{1/2}$. In the presence of the Hall effect \cite{2017ApJ...845...25P, 2020ApJ...902..142S}, a strong guide field does not change the dominance of the fastest-growing parallel mode but suppresses the wavelike structure of oblique modes, making the eigenfunctions asymmetric. These findings underscore the significance of including guide fields in tearing mode analyses, as they fundamentally alter the instability characteristics and contribute to the formation and dynamics of plasmoids in 3D MHD systems.

As discussed above, the tearing instability may be suppressed, or have its scaling laws severely modified, under some circumstances, either by viscous effects or the presence of velocity shears. Another stabilizing factor for tearing is the presence of a transverse, i.e. normal to the current sheet plane, magnetic field component. The stabilization effect of the transverse field $B_n$ becomes essential for $\xi \equiv {B_n}/{B_0} \geq S^{-3/4}$ \cite{1989SoPh..120...93S}, decreasing the growth rate $\gamma$ with increasing the value of $\xi$. Once $\xi \gg S^{-3/4}$, the instability is completely stabilized. This conclusion was, however, drew based on the stability of linearized incompressible MHD equations under the assumption of the same equilibrium as in the original work by Furth, Killeen, and Rosenbluth (1963).

In this work we demonstrate that the presence of transverse field modifies the initial magnetic shear profile. We provide the analytical solution of an initial Harris current sheet profile in the presence of transverse field, and perform the stability analysis of the incompressible MHD equations for the modified profile.

\section*{Harris Current Sheet in the Presence of a Transverse Field} \label{harris}

First, we analyze the effect of a transverse magnetic field on the evolution of the initial configuration of a one-dimensional (1D) Harris current sheet with thickness $a$. The initial magnetic field configuration is given by $B_x(t=0, z) = B_0 \tanh\left(\frac{z}{a}\right)$ for the field parallel to the current sheet, and $B_y(t=0, z) = B_0 \mathrm{sech}\left(\frac{z}{a}\right)$ for the guide field, with no initial flow ($U_x(t=0, z) = U_y(t=0, z) = 0$), and a non-zero, uniform transverse field ($B_z = \text{const}$). The evolution of the system along the $z$-direction is governed by the incompressible magnetohydrodynamic (MHD) equations:

\begin{equation} \label{eq:ux}
\frac{\partial U_x}{\partial t} = B_z \frac{\partial B_x}{\partial z} + \nu \frac{d^2 U_x}{dz^2}
\end{equation}

\begin{equation} \label{eq:uy}
\frac{\partial U_y}{\partial t} = B_z \frac{\partial B_y}{\partial z} + \nu \frac{\partial^2 U_y}{\partial z^2}
\end{equation}

\begin{equation} \label{eq:uz}
\frac{\partial U_z}{\partial t} = - \frac{\partial}{\partial z} \left[ p + \frac{1}{2} \left( B_x^2 + B_y^2 \right) \right]
\end{equation}

\begin{equation} \label{eq:bx}
\frac{\partial B_x}{\partial t} = B_z \frac{\partial U_x}{\partial z} + \eta \frac{\partial^2 B_x}{\partial z^2}
\end{equation}

\begin{equation} \label{eq:by}
\frac{\partial B_y}{\partial t} = B_z \frac{\partial U_y}{\partial z} + \eta \frac{\partial^2 B_y}{\partial z^2}
\end{equation}

\begin{equation} \label{eq:bz}
\frac{\partial B_z}{\partial t} = 0
\end{equation}
with the magnetic field expressed in terms of the Alfvén speed ($V_A \equiv B_0$), under the normalization used.

The chosen geometry and the condition of incompressibility enforce that the partial derivatives of the velocity component $U_z$ and the magnetic field component $B_z$ with respect to $z$ are zero ($\partial_z U_z = 0$ and $\partial_z B_z = 0$). Without loss of generality, $U_z = 0$ is initially assumed. The total pressure within the system is also found to remain uniform. From Equation~(\ref{eq:bz}), we infer that the transverse component of the magnetic field, $B_z$, likewise, remains constant over time allowing us to treat $B_z$ as a fixed parameter. Moreover, one can notice that Equations~(\ref{eq:ux}) and (\ref{eq:bx}) are separable from Equations~(\ref{eq:uy}) and (\ref{eq:by}).

The influence of the transverse field $B_z$ is manifested in Equations~(\ref{eq:ux}), (\ref{eq:uy}), (\ref{eq:bx}), and (\ref{eq:by}). Specifically, the term $B_z \partial_z B_x$ in Equation~(\ref{eq:ux}) induces a constant acceleration, resulting in a velocity shear in the X-direction, initially proportional to $B_z \text{sech}^2(z)$, provided $B_z$ is nonzero. In Equation~(\ref{eq:bx}), the shear in $U_x$ alters the profile of the current sheet through the term $B_z \partial_z U_x$. Furthermore, a non-zero $B_z$ impacts the Equations~(\ref{eq:uy}) and (\ref{eq:by}), where the non-uniform guide field $B_y$ generates another shear in the plane of the current sheet in the Y-direction via $B_z \partial_z B_y$, affecting the guide field $B_y$ profile due to $B_z \partial_z U_y$. The remaining non-ideal terms represent viscosity ($\nu$) in Equations~(\ref{eq:ux}) and (\ref{eq:uy}), and resistivity ($\eta$) in Equations~(\ref{eq:bx}) and (\ref{eq:by}).

By neglecting the effects of viscosity and resistivity, it becomes possible to analytically solve the system. Taking the time derivative of Equation~(\ref{eq:bx}) and substituting the time derivative of $U_x$ from Equation~(\ref{eq:ux}) leads us to a second-order differential equation for $B_x$:

\begin{equation}
\frac{\partial^2 B_x}{\partial t^2} = B_z^2 \frac{\partial^2 B_x}{\partial z^2}
\end{equation}

Given the initial Harris current sheet, described by a hyperbolic tangent, the solution to this second-order differential equation is:

\begin{equation} 
    B_x(t, z) = \frac{B_0}{2} \left[ \tanh{\left( \frac{z + B_z t}{a} \right)} + \tanh{\left( \frac{z - B_z t}{a} \right)} \right]
\end{equation}

It is easily verified that, at $t=0$, this profile corresponds to $B_0 \tanh{\left( \frac{z}{a} \right)}$. By introducing a relative measure of the transverse field with respect to the upstream field, i.e. $\xi \equiv B_z / B_0$, we have $B_z = ( B_z / B_0 ) B_0 = \xi V_A$.

The solution describes a current sheet of the thickness $a$ that develops a neutral layer of vanishing $B_x$ expanding around the midplane with the half-width increasing with time as $w = ({\xi}/{a}) V_A t$. The solution can be rewritten as:

\begin{equation} 
    B_x(t, z) = \frac{B_0}{2} \left[ \tanh{\left( \frac{z + \xi V_A t}{a} \right)} + \tanh{\left( \frac{z - \xi V_A t}{a} \right)} \right]
\end{equation}

From the magnetic field profile, we can deduce the solution for the shear $U_x$ by integrating the $z$-derivative of $B_x$ over time, as described by Equation~\ref{eq:ux}. The resultant expression for the $U_x$ shear is:

\begin{equation}
    U_x(t, z) = \frac{V_0}{2} \left[ \tanh{\left( \frac{z + \xi V_A t}{a} \right)} - \tanh{\left( \frac{z - \xi V_A t}{a} \right)} \right]
\end{equation}

It can be easily verified that at $t=0$, $U_x=0$. Upon examining this profile, it becomes apparent that its maximum occurs at $z=0$, which reveals the temporal increase in the amplitude of the $U_x$ shear as follows:

\begin{equation}
U_x(t, z=0) = V_0 \tanh{\left( \frac{\xi}{a} V_A t \right)}
\end{equation}

This equation implies that the amplitude of the $U_x$ shear initially increases from zero, growing linearly with time at a rate proportional to $\xi V_A / a$, and eventually saturates at the Alfvén speed beyond the saturation time $t_{\mathrm{sat}} \approx a / \xi V_A $.

Applying similar reasoning, we can determine the analytical solutions for $B_y$ and $U_y$, assuming negligible viscosity and resistivity. Taking the time derivative of Equation~\ref{eq:by} and inserting Equation~\ref{eq:uy} leads to a second-order differential equation:

\begin{equation}
\frac{\partial^2 B_y}{\partial t^2} = B_z^2 \frac{\partial^2 B_y}{\partial z^2}
\end{equation}

For the Harris current sheet model under consideration, the solution to this equation is:

\begin{equation} 
    B_y(t, z) = \frac{B_0}{2} \left[ \mathrm{sech}{\left( \frac{z + \xi V_A t}{a} \right)} + \mathrm{sech}{\left( \frac{z - \xi V_A t}{a} \right)} \right] + B_g,
\end{equation}
where $B_g$ is the uniform guide field component. Similarly, the solution for the $U_y$ shear profile:

\begin{equation} 
    U_y(t, z) = \frac{B_0}{2} \left[ \mathrm{sech}{\left( \frac{z + \xi V_A t}{a} \right)} - \mathrm{sech}{\left( \frac{z - \xi V_A t}{a} \right)} \right].
\end{equation}

The solutions for $B_y$ and $U_y$ hold under the condition that the guide field $B_g$ vanishes. Initially, the amplitude of this shear increases linearly from zero, reaching a saturation value of $V_A/2$. However, as the uniform component of the guide field $B_g$ increases, the amplitude of $U_y$ correspondingly decreases. Importantly, the evolution of $B_x$ and $U_x$ does not depend on the $y$-components of these fields, highlighting a distinct separation in the dynamics of the field components.

\section*{Numerical solutions of the current sheet evolution in the presence of a transverse field}

To validate our analytical predictions, we conducted numerical simulations of Equations~(\ref{eq:ux})-(\ref{eq:bz}) using the Dedalus framework (Dedalus Project: \href{https://dedalus-project.org/}{https://dedalus-project.org/}) on a setup consisting of two adjacent Harris current sheets within a periodic box. The initial current sheet thickness $a$ was set to unity. Our exploration included various values of the transverse magnetic field, $\xi$.

\begin{figure*}
    \centering
    \includegraphics[width=0.49\textwidth]{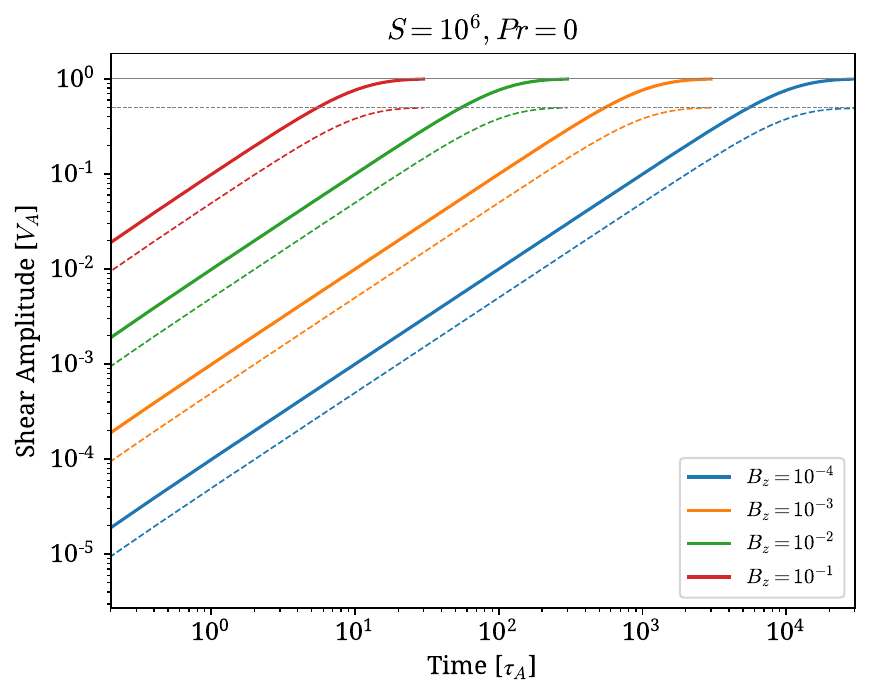}
    \caption{The growth of the shear amplitude $U_x$ (solid) and $U_y$ (dashed) as functions of the transverse field strength, $\xi$. The initial current sheet thickness was assumed to unity, $a=1$ and the null uniform guide field, $B_g = 0$. The horizontal gray lines indicate the saturation levels of $V_A$ (solid) and $V_A/2$ (dashed) for the shear velocities. The saturation time, $\tau_\mathrm{sat}$, scales proportionally with the strength of $\xi V_A$ ($t_\mathrm{sat} \sim a / \xi V_A$). These simulations were conducted with $S=10^6$ and $Pr=0$ ($\eta = 10^{-6}$ and $\nu = 0$, respectively).}
    \label{fig:harris-shear}
\end{figure*}

Figure~\ref{fig:harris-shear} illustrates the growth in the shear amplitudes of $U_x$ and $U_y$ within the current sheet setup across different $\xi$ values. Initially, both shear amplitudes demonstrate linear growth, progressing to a saturation level, in agreement with the analytical predictions. The time it takes to reach this saturation point scales with the magnitude of $\xi$. Notably, the $U_x$ shear reaches saturation at the Alfvén velocity ($V_A$), whereas the $U_y$ shear saturates at half the Alfvén velocity ($V_A/2$).

\begin{figure*}
    \centering
    \includegraphics[width=0.49\textwidth]{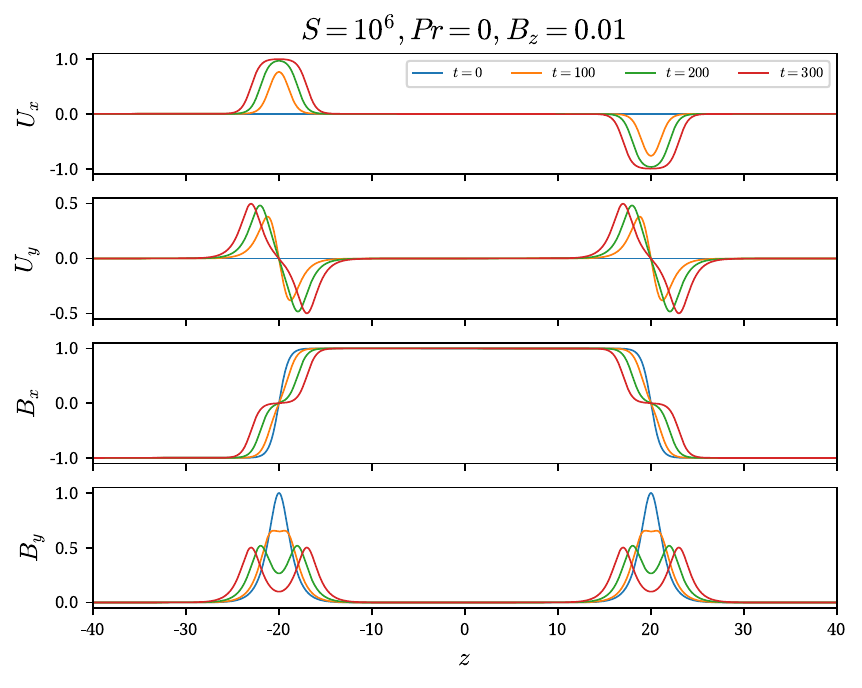}
    \caption{The field profiles at selected times ($t=0, 100, 200,$ and $300$) for a fixed transversal field, $B_z = 0.01$. The panels from top to bottom depict $U_x$, $U_y$, $B_x$, and $B_y$, respectively.}
    \label{fig:harris-shear-profiles}
\end{figure*}

Beyond the saturation phase, the current sheet undergoes expansion, indicated by the widening of the $B_x$ profile depicted in Figure~\ref{fig:harris-shear-profiles}. This expansion affects not only the $U_x$ shear region but also leads to the broadening of the profiles for $U_y$ and $B_y$, again, in agreement with the analytical predictions.

The analysis of the one-dimensional Harris current sheet in the presence of a transverse magnetic field reveals significant dynamics that are governed by the incompressible MHD equations. The study highlights how a constant transverse field $B_z$ influences the evolution of magnetic and velocity field profiles, inducing shears in both the $x$- and $y$-directions. These shears evolve due to the coupling between velocity and magnetic field components, as mediated by $B_z$. The analytical solutions derived demonstrate that the magnetic field profile $B_x$ and velocity profile $U_x$ both split and propagate away from the midplane, leading to the development of dynamic shears that increase linearly in amplitude and eventually saturate. This saturation occurs at the Alfvén speed for $U_x$ and at half the Alfvén speed for $U_y$, dependent on the magnitude of $B_z$. Furthermore, the evolution exhibits a distinct separation in the dynamics of the $x$- and $y$-components, underscoring the robust separability in their behaviors. Numerical simulations confirm these analytical findings, illustrating the key role of the transverse field in shaping the temporal and spatial dynamics of the Harris current sheet.

\section*{Tearing Instability Analysis in the Quasi-Steady Expanding Current Sheet Limit}\label{stability}

In the previous section, we established that the presence of a transverse field disrupts the equilibrium of the 1D Harris current sheet, altering its profile and creating an expanding neutral layer. If $B_z$ is sufficiently small, this expansion may be slow enough to allow the development of tearing instability, which can be studied in a quasi-steady current sheet scenario. 

In this section, we will perform a stability analysis using the current sheet profile with a neutral layer as defined by the solution found in the previous section. We perform such analysis using different profiles, that correspond to different evolutionary stages of the current sheet. Apart from the $B_x$ profile, we will also consider a corresponding solution for $U_x$. We parameterize the strength of the transverse field by $\xi$ as defined earlier, and the half-width of the expanding neutral layer by $w = \xi V_A t$. Assuming, without loss of generality, that $B_0 = 1$ and $V_A = 1$.

The linearized equations are derived from the incompressible MHD equations in a traditional manner, assuming a 2D setup in X and Z coordinates. To eliminate the dependency on pressure, we take a curl of the momentum equation. We assume that the perturbations take the form $\hat{q}(t, x, z) = q(z) \exp{\left( - \gamma t + i k x \right)}$. Furthermore, we substitute the x components of velocity and magnetic field Fourier amplitudes using the incompressibility condition, i.e., $b_x = i \partial_z b_z / k$ and $u_x = i \partial_z u_z / k$. This procedure leads to the following linearized independent equations:

\begin{equation} \label{eq:duz}
\begin{aligned}
    \gamma \left( u_z'' - k^2 u_z \right) = i k \left[ F \left( b_z'' - k^2 b_z \right) - F'' b_z - G \left( u_z''- k^2 u_z \right) + G'' u_z \right] \\ \textcolor{blue}{+ \xi \left( b_z''' - k^2 b_z' \right)} + P_r S^{-1} \left( u_z'''' - 2 k^2 u_z'' + k^4 u_z \right),
\end{aligned}
\end{equation}

\begin{equation} \label{eq:dbz}
\begin{aligned}
    \gamma b_z = i k \left( F u_z - G b_z \right) \textcolor{blue}{+ \xi u_z'} + S^{-1} \left( b_z'' - k^2 b_z \right),
\end{aligned}
\end{equation}
where $\xi \equiv B_t / B_0$ is the relative strength of the transverse field, and $'$, $''$, $'''$, and $''''$ are the first, second, third, and fourth-order derivatives with respect to $z$, respectively. The functions $F(z) = \frac{1}{2} \left[ \tanh \left(\frac{z + w}{a}\right) + \tanh \left(\frac{z - w}{a}\right) \right]$ and $G(z) = \frac{1}{2} \left[ \tanh \left(\frac{z + w}{a}\right) - \tanh \left(\frac{z - w}{a}\right) \right]$ represent the analytical solutions for $B_x$ and $U_x$, respectively, at time $t$, corresponding to $w = \xi V_A t$.  Here, $L$ represents the assumed unit of length. Furthermore, we define two Alfvén time scales: $\tau_A = a / V_A$, related to the thickness of the current sheet, and $t_A = L / V_A$, related to the characteristic scale $L$. For most cases presented here, we assume $L = a = 1$, resulting in $\tau_A$ and $t_A$ being equal. The linearized equations (\ref{eq:duz})-(\ref{eq:dbz}) resemble those derived in previous studies on tearing instability \cite[see, e.g., ][]{2015ApJ...806..131L, 2018ApJ...859...83S}. The distinction arises from the presence of terms related to the shear, represented by $G$ and $G''$, and terms containing the constant transverse field, $\xi$.

We perform the stability analysis of the Equations~(\ref{eq:duz})-(\ref{eq:dbz}) using the \textsc{Psecas} framework \cite{2019MNRAS.485..908B}, which offers a robust tool for studying the stability of various physical systems by facilitating the solution of linear stability problems. It is particularly adapted for handling differential equations that describe perturbations in continuous media, such as those found in fluid dynamics and plasma physics. The framework employs spectral methods, which are advantageous for their accuracy and efficiency in resolving complex waveforms in stability analyses. By decomposing the problem space into basis functions, \textsc{Psecas} efficiently transforms the differential equations into a matrix eigenvalue problem. This allows for the computation of growth rates and eigenmodes, essential for understanding the stability characteristics of the system. The flexibility of \textsc{Psecas} in selecting appropriate basis functions and boundary conditions makes it a versatile tool in the theoretical investigation of instabilities, enhancing both the depth and scope of analysis possible within computational research environments.

\begin{figure*}
    \centering
    \includegraphics[width=0.49\textwidth]{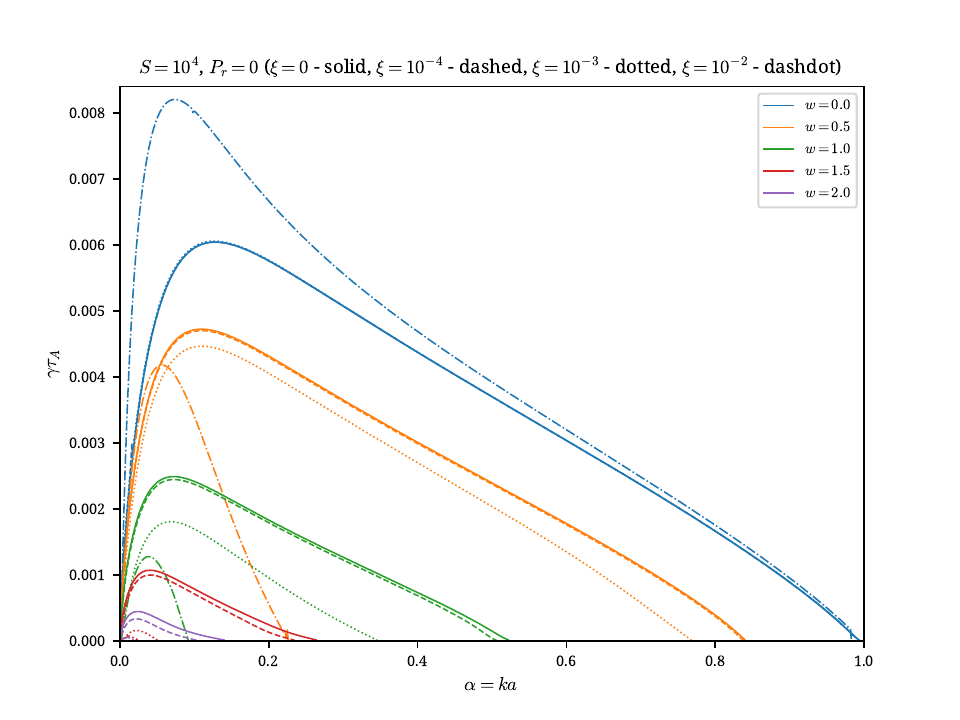}
    \includegraphics[width=0.49\textwidth]{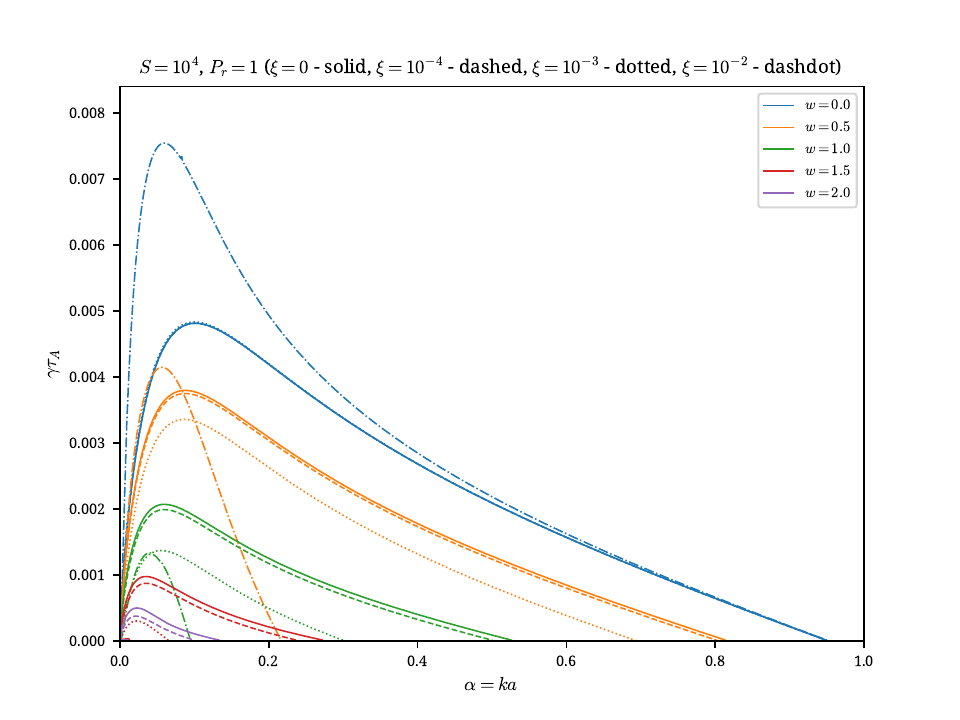}
    \caption{The dispersion relations for the case of Lundquist number $ S=10^4 $ are shown for several values of $ w = 0, 0.5, 1, 1.5, $ and $ 2 $ (colors), and strengths of the transverse field $ \xi = 0, 10^{-4}, 10^{-3}, $ and $ 10^{-2} $ (solid, dashed, dotted, and dashdot line styles, respectively). The upper and lower plots correspond to cases with Prandtl numbers 0 and 1, respectively. For this analysis we assumed here $L=a=1$.}
    \label{fig:dispersions}
\end{figure*}

Figure~\ref{fig:dispersions} shows dispersion relations for the equilibrium profiles of $B_x$ and $U_x$, described by functions $F$ and $G$, respectively, with several values of $w$. The solid lines correspond to $\xi = 0$ and are derived by neglecting the terms related to $\xi$ in Equations~(\ref{eq:duz})-(\ref{eq:dbz}). The system's response is solely due to the equilibrium profiles.

Two regimes of tearing instability can be recognized: the constant-$\psi$ regime at large $\alpha$, where the growth rate $\gamma$ decreases with the wavenumber, and the non-constant-$\psi$ regime at small $\alpha$, with $\gamma$ increasing with $\alpha$. The existence of these two regimes results in a maximum growth rate $\gamma_{\mathrm{max}}$ and corresponding wavenumber $\alpha_{\mathrm{max}}$. The figure shows that the dispersion relations decrease in terms of the growth rate $\gamma$ and shift their maximum to longer wavelengths with increasing $w$. For $w = 2$, the dispersion relation is characterized by growth rates smaller than $5 \times 10^{-4}$ with wavelengths $\alpha = ka$ smaller than 0.2. Including the viscous terms in Equation~(\ref{eq:duz}) results in a further decrease of the growth rates $\gamma$ (compare upper and lower panels in Fig.~\ref{fig:dispersions}).

The dispersion relations calculated by taking into account the terms with $\xi$ in Equations~(\ref{eq:duz})-(\ref{eq:dbz}) are plotted with dashed, dotted, and dash-dotted lines for $\xi = 10^{-4}$, $10^{-3}$, and $10^{-2}$, respectively. Small values of $\xi \ll 10^{-2}$, result in a gradual decrease of the growth rates at all wavelengths. For $w = 0$ and $\xi = 10^{-2}$, we observe a significant increase of the growth rate, especially in the region of $\gamma_{\mathrm{max}}$. However, as the system evolves and the neutral layer width ($w$) increases, the growth rate in the non-constant-$\psi$ regime drops, and most of the wavelengths in the constant-$\psi$ regime are stabilized.

These results demonstrate the significant impact of the $\xi$-related terms on tearing mode instability, enhancing the growth rates for the classical profile of the Harris current sheet ($w = 0$) with a sufficiently large transverse field $\xi=10^{-2}$, but quickly stabilizing most of the tearing instability modes once the equilibrium profiles expand due to the action of a non-zero $\xi$. It is important to note that the value of $w = 1$ corresponds to the half-width of the neutral layer being equal to the initial Harris current sheet thickness $a = 1$. Therefore, once the action of the transverse field expands the neutral layer to more than twice the thickness of the initial current sheet, the expected result is the stabilization of essentially all tearing instability modes.

\begin{figure*}
    \centering
    \includegraphics[width=\textwidth]{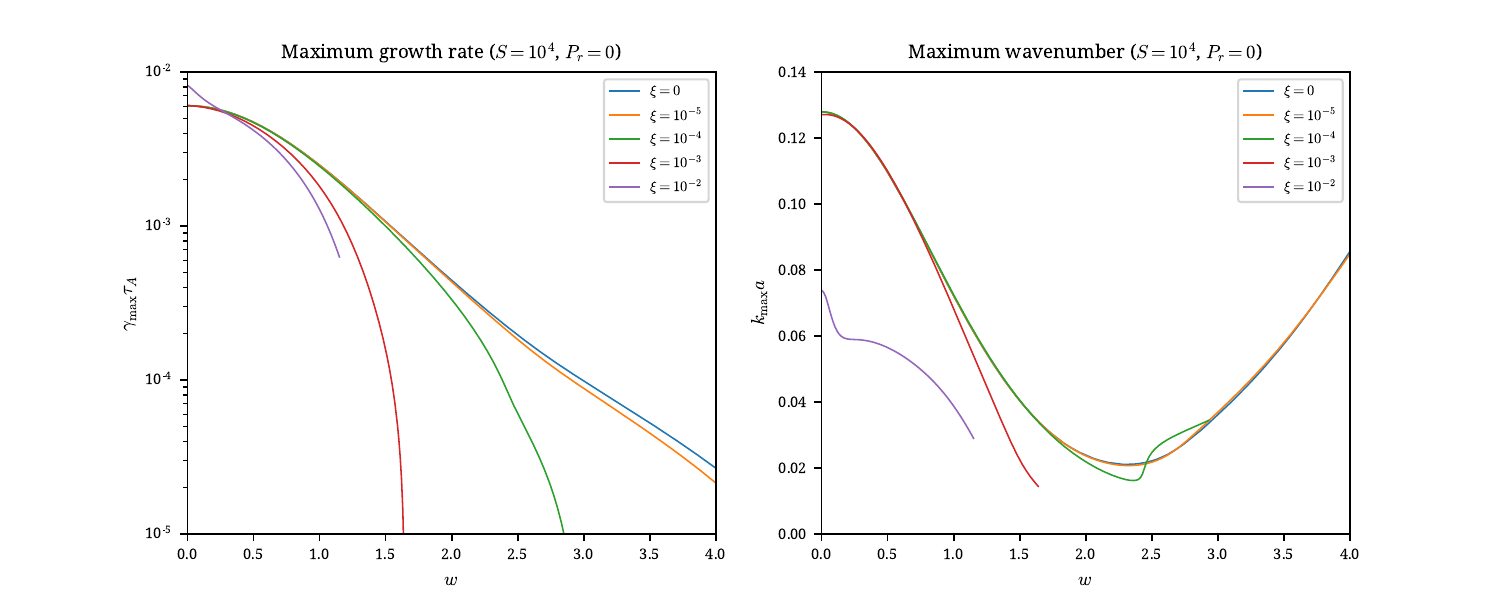}
    \includegraphics[width=\textwidth]{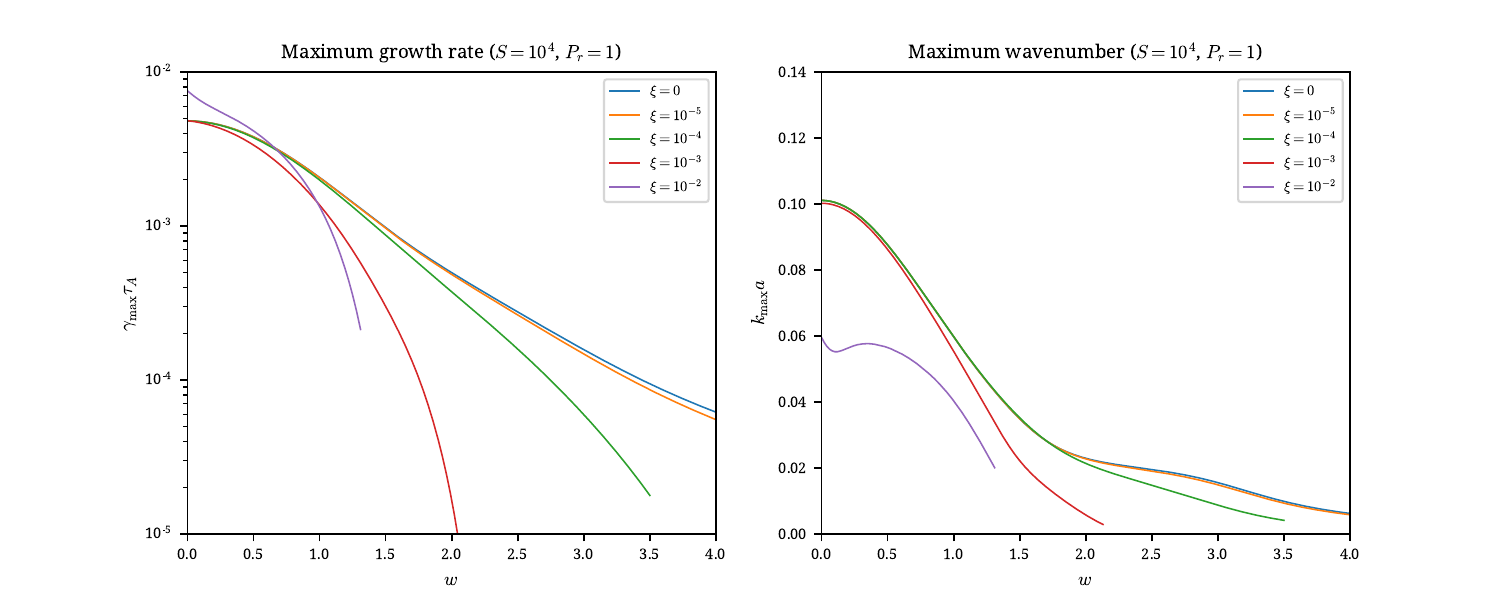}
    \caption{The dependence of maximum growth rate $\gamma_\mathrm{max}$ (left) and corresponding wavenumber $\alpha_\mathrm{max}$ (right) on the neutral layer width $w$ for different strength of transverse field $\xi$ ($0$, $10^{-5}$, $10^{-4}$, $10^{-3}$, and $10^{-2}$). The case of $S=10^4$ for $P_r = 0$ and $P_r = 1$ are shown in the upper and lower panels, respectively. For this analysis we assumed here $L=a=1$.}
    \label{fig:w_dep}
\end{figure*}

Now, focusing on the fastest growing mode, Figure~\ref{fig:w_dep} displays the maximum growth rate, $\gamma_\mathrm{max}$, and the corresponding wavenumber, $\alpha_\mathrm{max}$, in the left and right panels, respectively, dependence on the half-width of the neutral layer $w$ for cases with varying magnitudes of the transverse field, $\xi$. It is noticeable that for very weak transverse fields, $\xi < 10^{-4}$, the maximum growth rate decays nearly exponentially with increasing $w$. Once the transverse field strengthens, i.e., $\xi \ge 10^{-4}$, the maximum growth rate drops faster than the exponential rate, eventually stabilizing completely at a certain thickness of the neutral layer. 

In the inviscid case shown in the upper panel, the estimated stable half-widths are approximately $w \sim 2.6$, $\sim 1.7$, and $\sim 1.2$ for $\xi = 10^{-4}$, $10^{-3}$, and $10^{-2}$, respectively. A similar behavior is observed in the viscous case (lower panel), with stable half-widths slightly larger. We anticipate that the action of the viscous terms in Equation~(\ref{eq:duz}) decreases the gradients in the $u_z$ profiles, which somewhat diminishes the impact of the $\xi$-related term in Equation~(\ref{eq:dbz}).

\begin{figure*}
    \centering
    \includegraphics[width=\textwidth]{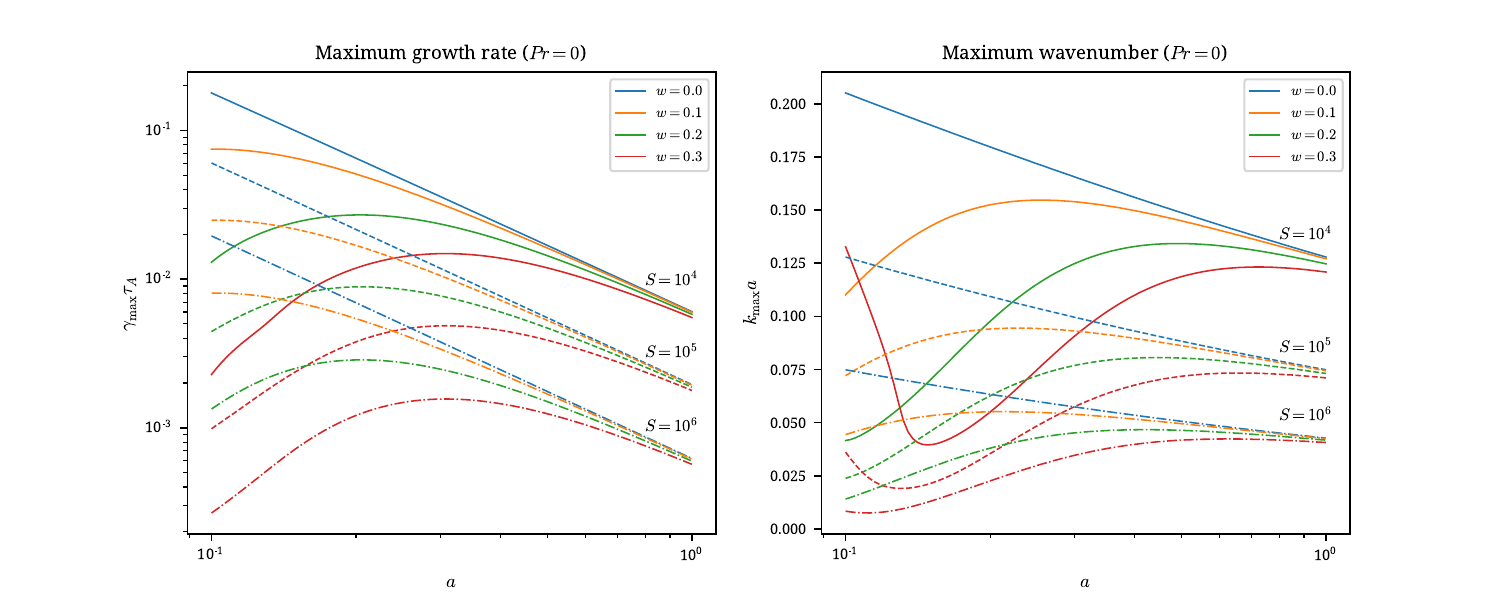}
    \includegraphics[width=\textwidth]{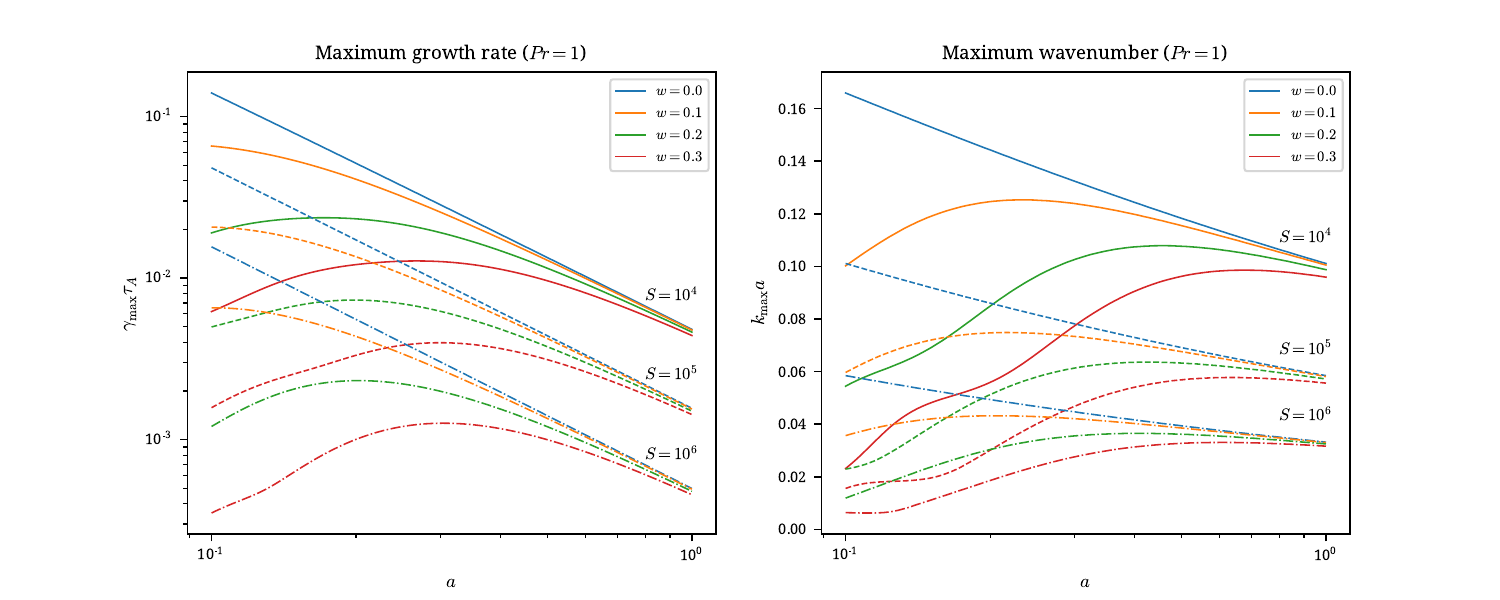}
    \caption{The dependence of maximum growth rate $\gamma_\mathrm{max}$ (left) and corresponding wavenumber $\alpha_\mathrm{max}$ (right) on the initial current sheet thickness $a$ for different neutral layer widths $w$ ($0$, $0.1$, $0.2$, and $0.3$) and Lundquist numbers ($10^4$, $10^5$, and $10^6$). The case of $Pr = 0$ and $Pr = 1$ are shown in the upper and lower panels, respectively.}
    \label{fig:a_dep}
\end{figure*}

Finally, we study the dependence on the initial current sheet thickness, $a$. In the classical tearing mode analysis the maximum growth rate increases with decreasing current sheet thickness as $\gamma_\mathrm{max} \propto (a/L)^{-3/2}$. As shown in the left panels of Figure~\ref{fig:a_dep}, this behaviour is recovered only for $w = 0$. Once the neutral layer increases, the behaviour of $\gamma_\mathrm{max}$ with $a$ changes drastically. We show that in the presence of the expanding neutral layer this relation does not hold anymore, mostly due to the changing profile of the equilibrium field, and not the action of $\xi$ related terms. The saturation time $t_\mathrm{sat} \propto a$, therefore, for smaller $a$ the tearing is stabilized faster. Indeed, as seen in Figure~\ref{fig:a_dep}, already for 10 times thinner current sheet the growth rate decays with decreasing $a$ for $w > 0.1$. At the same time, the corresponding wavenumber $k_\mathrm{max}$ tends to larger scales, as shown in the right panels of Figure~\ref{fig:a_dep}.

\section*{Direct Numerical Simulations of Tearing Mode in an Expanding Current Sheet}\label{simulations}

The stability analysis of the previous section, performed for fixed equilibrium profiles, indicates that the tearing mode instability should be suppressed for a sufficiently thick neutral layer. To verify if this conclusion holds for time-varying profiles of magnetic and velocity shears, we conducted direct numerical simulations of the initial Harris current sheet, represented by a hyperbolic tangent profile of the $x$-component of the magnetic field and varying transverse field strength. The numerical solution of the incompressible MHD equations in 2D was obtained using the Dedalus framework (Dedalus Project: \href{https://dedalus-project.org/}{https://dedalus-project.org/}).

\begin{figure*}
    \centering
    \subfloat[]{\includegraphics[width=0.49\textwidth]{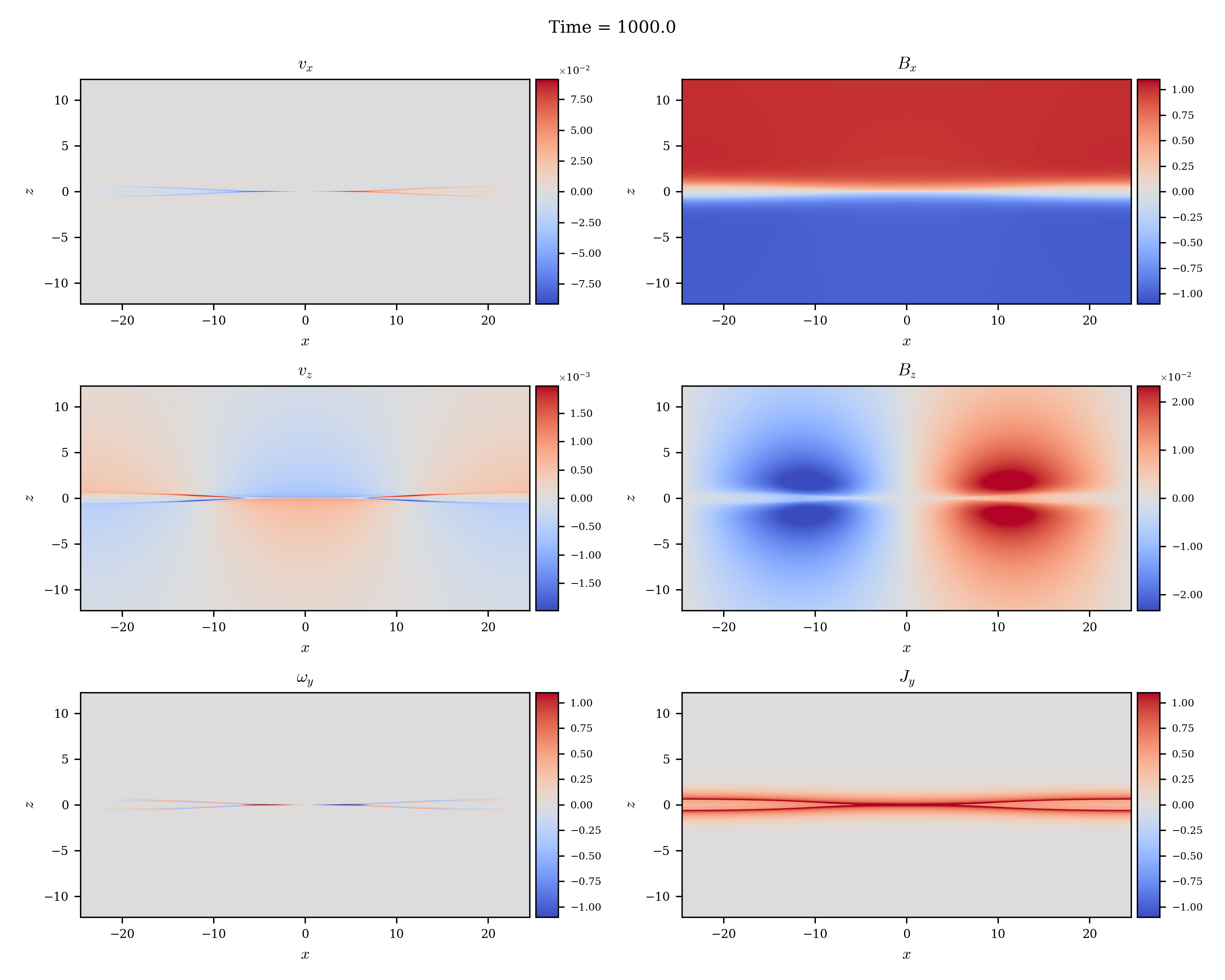}}
    \subfloat[]{\includegraphics[width=0.49\textwidth]{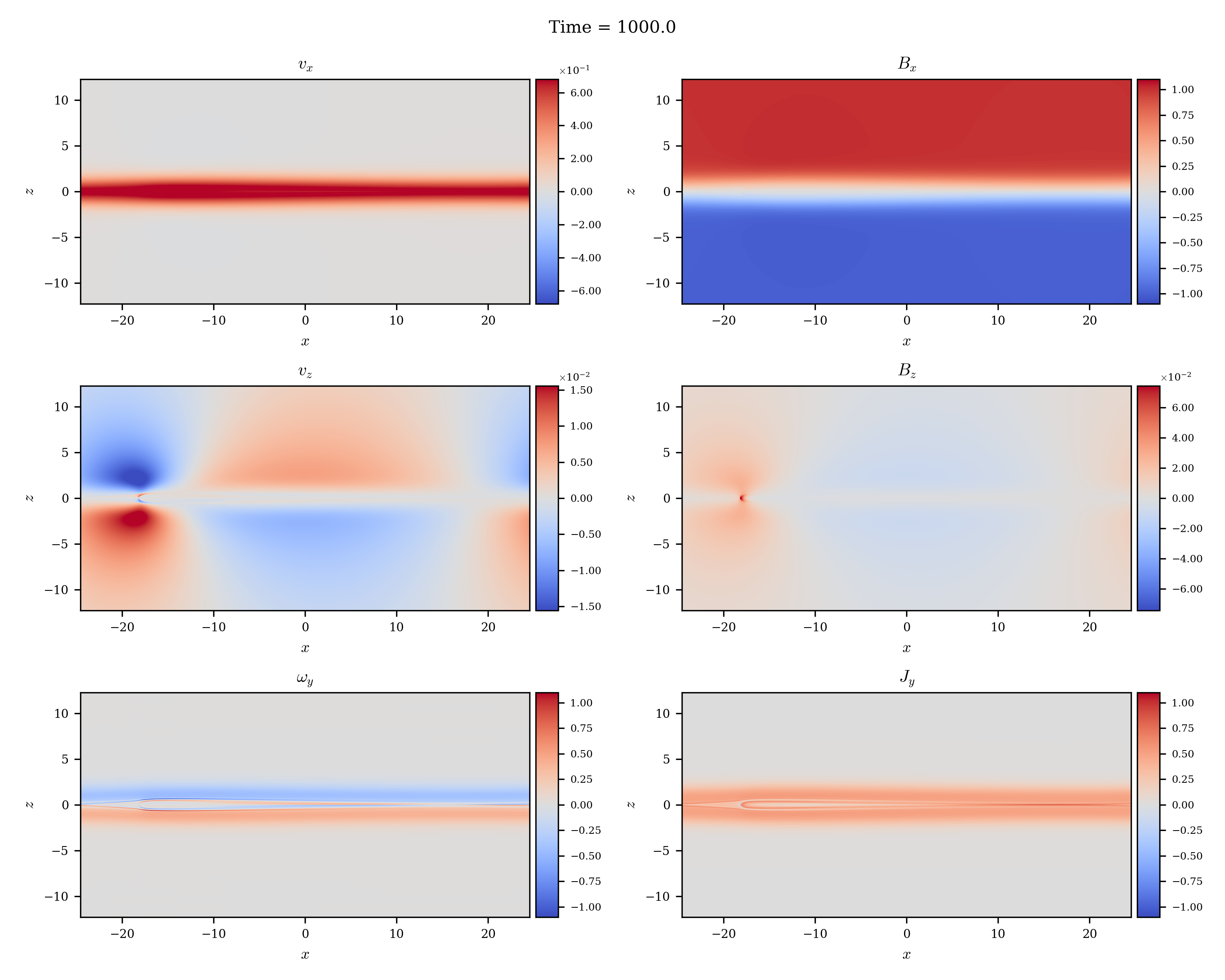}} \\
    \subfloat[]{\includegraphics[width=0.49\textwidth]{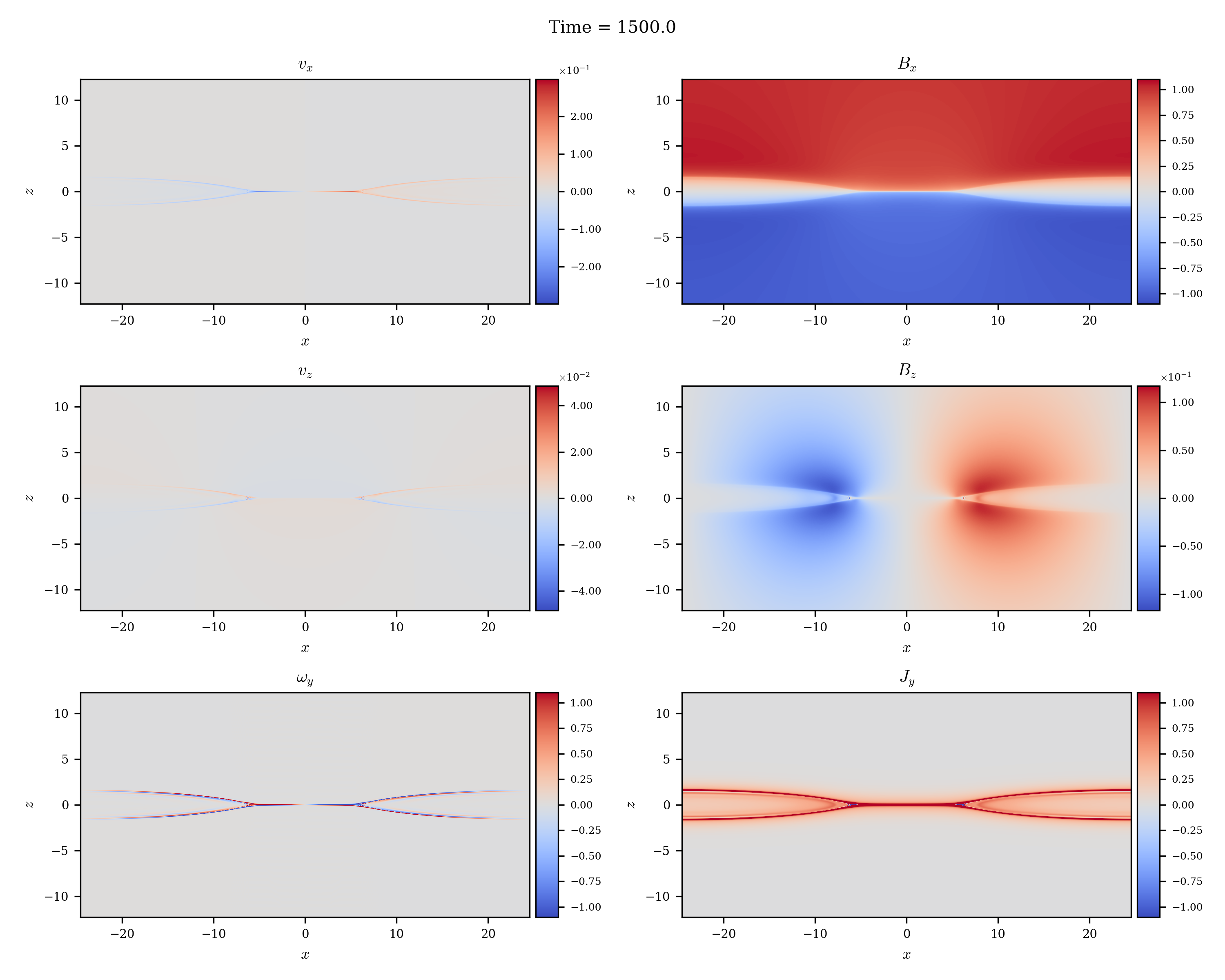}}
    \subfloat[]{\includegraphics[width=0.49\textwidth]{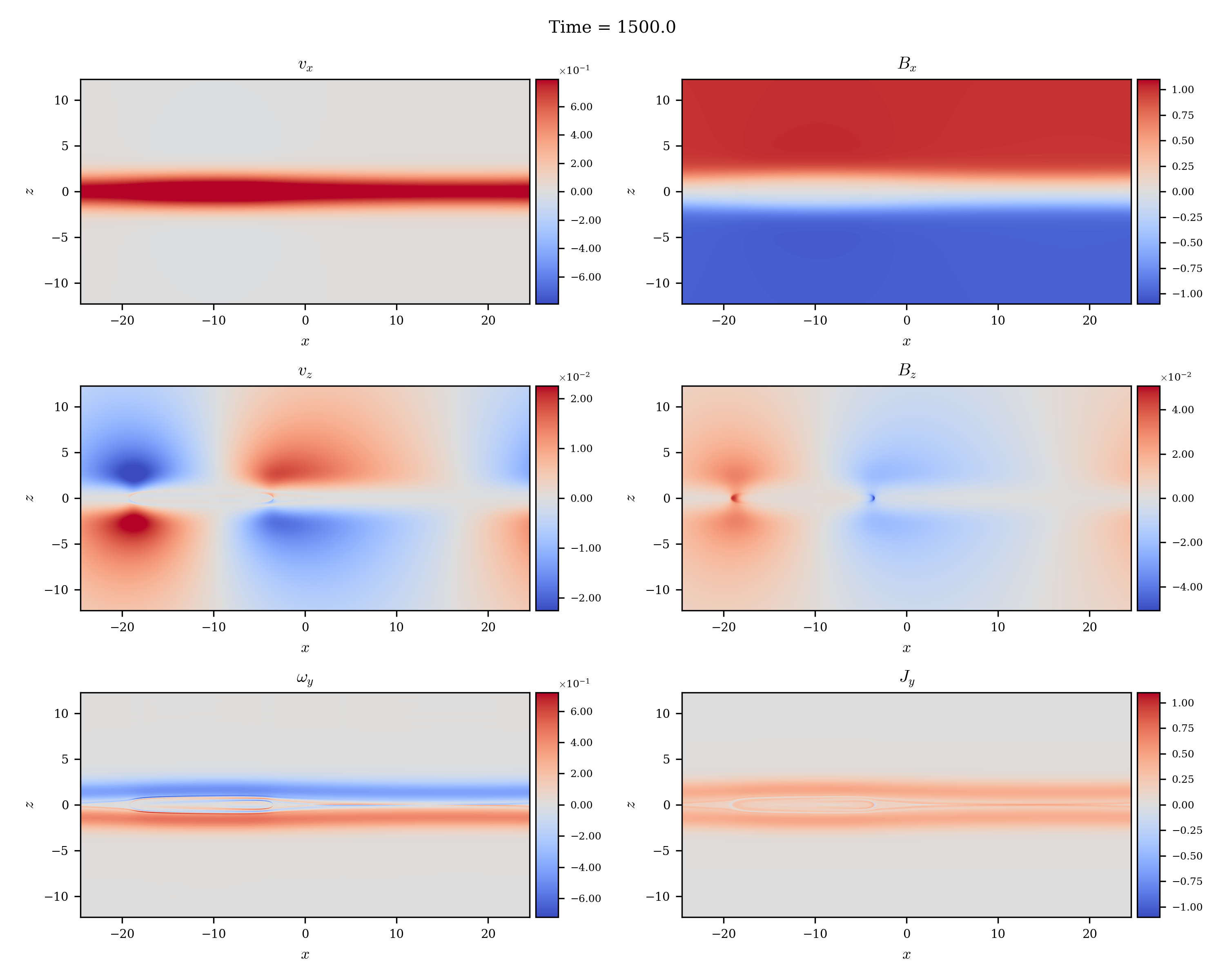}}
    \caption{Field profiles in simulations of the initial Harris current sheet equilibrium with two values of transverse field, $\xi = 0$ (subfigures a and c) and $\xi = 10^{-3}$ (subfigures b and d) at two different moments, $t=1000$ (subfigures a and b) and $t=1500$ (subfigures c and d). Each subfigure contains colormaps of $v_x$, $v_z$, $\omega_y$ (left panels from top to bottom, respectively), and $B_x$, $B_z$, $j_y$ (right panels from top to bottom, respectively). The simulations were conducted with a Lundquist number $S = 10^4$, magnetic Prandtl number $Pr_m = 0$, and initial current sheet thickness $a = 1$.}
    \label{fig:simulations}
\end{figure*}

Figure~\ref{fig:simulations} shows examples of simulations performed with a Lundquist number $S = 10^4$ and a magnetic Prandtl number $Pr_m = 0$, for the initial Harris current sheet of thickness $a = 1$ and different strengths of the initial transverse field, $B_z = 0$ (subfigures a and c) and $B_z = 10^{-3}$ (subfigures b and d). The amplitude of the initial magnetic field perturbation was set to $10^{-4}$. The velocity components and vorticity are shown in the left panels of each subfigure, while the magnetic field components and current density are shown in the right panels. The growth rate of the most unstable tearing mode for this case is $\gamma_\mathrm{max} \approx 0.006$, resulting in an instability time scale of $t_\mathrm{tearing} \approx 165 \tau_a$, where $\tau_a$ is the current sheet crossing time. Two moments are depicted: $t = 1000$ (subfigures a and b) and $t = 1500$ (subfigures c and d). At these times, the tearing mode should grow by factors of approximately 420 and 8,600, respectively. For comparison, the estimated stabilization time $t_\mathrm{sat}$ due to the presence of the transverse field for the case of $\xi = 10^{-3}$ is around 1000, several times longer than the instability growth time.

In subfigures a and c of Figure~\ref{fig:simulations}, one can recognize the typical evolution of tearing mode instability in the case of a null transverse field: thickening of the current sheet in the center, accompanied by outflows from the center in both directions, resulting in the creation of two plasmoids on both sides. In the case of a non-zero but weak transverse field, shown in subfigures b and d, we observe the formation of an expanding neutral layer with a strong velocity shear, as predicted by the analytical solution of the 1D Harris current sheet evolution. Additionally, we see a much weaker current density with no characteristic plasmoid formation, separated by a thin current sheet.

In conclusion, the 2D numerical simulations of the initial Harris current sheet with a non-zero transverse field confirm the prediction from the linear stability analysis regarding the stabilization effect of the transverse field on tearing mode instability. The presence of a transverse field indeed suppresses the tearing mode instability, leading to the formation of a stable, expanding neutral layer as opposed to the plasmoid-dominated structure seen in the absence of a transverse field.

\section*{Conclusions}

In this work, the impact of a transverse magnetic field on the evolution of a Harris current sheet for magnetic reconnection is thoroughly investigated, including its impact on the tearing/plasmoid instability. The showed that the presence of the transverse field $B_z$ modifies the initial shear profile of the magnetic field, as demonstrated through both analytical and numerical methods. The derived analytical solutions for the magnetic and velocity fields show that $B_z$ induces, and enhances through time, a velocity shear in the $x$-direction, which eventually saturates at the Alfvén speed. This transverse field also affects the $y$-direction shear, although it saturates at half the Alfvén speed. In a time evolution analysis, we showed that this increased shear induces the broadening of the current sheet, resulting in the formation of a neutral layer whose width increases over time. This effect becomes more important for stronger transverse fields, 
Numerical simulations validate these analytical predictions, indicating a clear alignment between theoretical and computational results. 

Major consequences of the time-evolution described above is seen in the tearing/plasmoid instability. The stability analysis of the current sheet, taking into account the transverse field, indicates significant modifications on the modes of tearing instability. The analysis reveals that for small values of $B_z$ the growth rates of tearing instabilities are reduced, leading to stabilization as the neutral layer expands. For strong transverse fields the growth rates initially increase but are quickly reduced as the neutral layer broadens. This demonstrates that, despite what has been previously thought, the combined effect of a strong transverse field in tearing, and the time evolution of the current sheet, actually stabilizes the tearing modes by expanding the neutral layer beyond a critical thickness. 

Direct numerical simulations of the tearing instability with varying transverse field strengths further support the theoretical predictions, demonstrating the stabilizing effect of the transverse field. The direct numerical simulations show that, in the absence of a transverse field, the tearing mode instability leads to the formation of plasmoids and a thickened current sheet. However, with a non-zero transverse field, an expanding neutral layer with strong velocity shear develops, and the characteristic plasmoid formation is suppressed. The stabilization effect is clearly observed, as the transverse field prevents the typical tearing mode evolution and results in a stable, expanding neutral layer. These results validate the theoretical stability analysis presented in this work, confirming that a sufficiently strong transverse field can effectively quench the tearing mode instability.

Our study emphasizes the role of the transverse magnetic field in shaping the evolution and stability of current sheets, highlighting the intricate interplay between magnetic field configurations and plasma instabilities. These findings may change considerable the understanding of magnetic reconnection and turbulence in plasma physics, with potential implications for space and astrophysical phenomena, given that one of the fast mechanisms of magnetic reconnection has been shown here to be quickly quenched in real systems.

Stabilization effects of transverse fields has been already pointed by Somov and Verneta \cite{1988SoPh..117...89S, 1989SoPh..120...93S, 1993SSRv...65..253S}. In their work, Somov and Verneta demonstrated a significant stabilizing influence of a transverse magnetic field within the magnetohydrodynamics (MHD) framework, highlighting how a small transverse field can suppress the tearing mode instability in current layers. They also explained the discrepancies found in earlier studies, which had yielded negative results regarding the stabilizing effects. The current study advances this understanding by providing detailed analytical solutions for the evolution of magnetic and velocity shears in the presence of a transverse field and confirming these theoretical predictions through direct numerical simulations using the Dedalus framework. The simulations show that the transverse field indeed leads to the formation of an expanding neutral layer and suppresses the characteristic plasmoid formation associated with tearing modes, thus corroborating the theoretical insights provided by Somov and Verneta and extending them to time-varying profiles in a dynamically evolving current sheet. This alignment between analytical and numerical results offers a robust confirmation of the stabilizing effects of transverse fields on tearing mode instabilities in reconnecting current layers.

\bibliography{bibliography}




\section*{Acknowledgments}

G.K. and D.F-G acknowledges support from FAPESP (grants 2013/10559-5, 2021/02120-0, 2021/06502-4, and 2022/03972-2). The simulations presented in this work were performed using the clusters of the Group of Theoretical Astrophysics at EACH-USP (Hydra HPC), which was acquired with support from FAPESP (grants 2013/04073-2 and 2022/03972-2).

\section*{Author contribution}

G.K. derived the analytical solution of 1D MHD for initial Harris current sheet and testes this solution numerically. D.F.G. performed the linearization of incompressible MHD equations taking into account the viscosity, resistivity and transverse field related terms. G.K. implemented these equations in the Psecas framework and performed the calculations. G.K. performed 1D and 2D simulations of current sheet in the presence of transverse field using the Dedalus framework. Both authors analyzed and interpreted the obtained dispersion relations for different profiles of the equilibrium, and Lundquist and Prandtl numbers, as well as, the direct numerical simulations.

\section*{Competing interests}

The authors declare no competing financial interest.

\end{document}